# Learning Entrepreneurhsip with Serious Games - A Classroom Approach

Fernando L. F. Almeida
Informatics Engineering Department
Center of Innovation, Technology and Entrepreneurship
Faculty of Engineering of University of Porto, ISPGaya and INESC TEC
Porto, Portugal
falmeida@ispgaya.pt

*Abstract -* **The use of educational games for pedagogical practice can provide new conceptions of teaching-learning in an interactive environment stimulating the acquisition of new knowledge. The so-called serious games are focused on the goal of transmitting educational content or training to the user. In the context of entrepreneurship, serious games appear to have greater importance due to the multidisciplinary of the knowledge needed. Therefore, we propose the adoption of the *Entrexplorer game* in the context of a university classroom. The game is a cloud-based serious game about the theme of entrepreneurship where users can access learning contents that will assist them in the acquisition of entrepreneurial skills. The organization of the game in eight levels with six additional floors let students learn the different dimensions of an entrepreneurship project while progressing during the gameplay.**

## I. INTRODUCTION

With the advent of globalization, which is based on the pillars of knowledge and technology, education is seen as the greatest resource available to face the challenges of this dynamic world. In fact, our current process of economic and social development is based on the principles of a good education. Creative entrepreneurs emerge as the fuel for a creative economy. Jenkins (2016) defined a creative entrepreneur as a person who believes in building intrinsic wealth in self and others versus acquiring capitalistic wealth. Creative entrepreneurialism has three distinctive elements: creativity, collaboration, and entrepreneurship.

Serious games in the context of a university classroom can be used to foment the characteristics of a creative entrepreneurialism. In fact, the potential of serious games in education is widely recognized, and their adoption is significant in particular in children instruction (Bellotti et al., 2010). Bushnell referred by Tack (2013) states that serious games are the most attractive path for the future of education. However, the deployment rate of SGs in higher education and their proper insertion in meaningful curricula is still quite low.

In fact, the use of games by students brings additional challenges regarding the design of games and their adoption in different learning, academic and interdisciplinary contexts (Almeida et al., 2015). Some of the opponents support the idea that the effectiveness of the method can hardly be evaluated. They also think that their inclusion in corporate and formal training programs requires a lot of resources connected not only with the provision of sufficient amount of funding, but also with recruitment of teachers and trainers possessing necessary knowledge and skills (Karner & Härtel, 2011). On the other side, there are also some arguments in favor of the application of serious game based learning approach, namely in terms of enhancing productivity, enhancing the level of appreciation of the learners' role, testing competencies, evaluation, training and best practices exchanges.

The remainder of this paper is organized as follows: In Section II, we make a brief literature review in the field of serious games and entrepreneurship. Then, Section IV presents the mission and structure of *Entrexplorer* project. Finally the conclusion is drawn in Section IV.

## II. LITERATURE REVIEW

### A. Serious games in education

Games have become a major recreational activity, and they have also become increasingly sophisticated and celebrated as a cultural form; they have shaken up the world of entertainment, and they have entered into educational debates and practices. There are different possibilities to distinguish computer games, one of them is to divide them up into the following categories:

- o Casual games, which are developed purely as entertainment activity, and thus the learning outcome are not intentionally foreseen;
- o Advertising games, which are identified as tools designed and delivered as promotion and marketing

- of products, services, new coming movie or TV series;
- Serious games, which are especially designed to improve some learning aspects and players expect the learning process.

Breuer and Bente (2010) propose nine categories for classifying serious games: (i) platform (e.g., personal computer, android, etc.); (ii) subject matter (e.g., energy, entrepreneurship, etc.); (iii) learning goals (e.g., language skills, historical facts, etc.); (iv) learning principles (observational learning, trial and error, etc.); (v) target audience (e.g., higher education, pre-schoolers, etc.); (vi) interaction mode(s) (e.g., single player, multiplayer, etc.); (vii) application area (e.g., academic education, private use, etc.); (viii) controls/interfaces (e.g., gamepad controlled, mouse & keyboard, etc.); (ix) common gaming labels (e.g., puzzle, simulation, etc.).

Learning games rise especially since computers and the internet makes a lot of interesting and exciting games possible. They are in use in any devices and in any environment. Not only because of intentional learning. Another point is the possibility to motivate the user to handle with topics, which are boring for them, if they had to learn that content in a traditional way or with traditional media. The aim of using a game is to foster interest and motivate the target group to handle with the topic of the learning game. The game is only the means of transportation for the pedagogical content in the game. So the concept of the game - the idea behind the game has to be adjusted to the target group. The tasks for the users, the graphics and animation and so on, have to be adjusted to the target group. The content is exchangeable.

Serious games are games with a purpose beyond entertainment and deal with issues related to learning, health and politics, among others. Michael & Chen (2006) define serious games as "games that do not have entertainment, enjoyment, or fun as their primary purpose" or "a serious game is a game in which education is the primary goal, rather than entertainment".

Giessens (2015) states that is crucial for a serious game to find a balance between the ludative element that is present in all games, and didactical or pedagogical goals that should exist in a non-intrusive manner. At the same time, it is important to notice that enjoying the game does not automatically mean learning success. It is important to encourage student's motivation to learn about a subject, but engagement with content is essential in this process (Iten and Petko, 2016).

Cruz (2008) considers that in order to be effective a serious game should ensure the four key characteristics of a successful game: challenge, skills exercise, competition and a sense of progress. The serious games are designed with the intention of improving specific aspects of learning and their users seek this activity based on these expectations. The serious games are used for training of emergency services, military training, organizational education, medical care and many other sectors of society. They can also be found at all educational levels and in all schools and universities all over the world. The act of playing is an essential ingredient in a serious game, since it is an important contribution to the maturation, learning and human development (Cruz, 2008).

Mouaheb et al. (2012) clearly identify three major educational advantages of adopting serious games in classrooms:

- Offer an intense interaction that generates real cognitive or socio-cognitive conflicts, providing a solid construction of knowledge;
- Provide an autonomy in the learning process following a strong meta-cognitive activity;
- Potentiate an eventual transfer of acquired skills.

*B. Entrepreneurship in Higher Education*

Entrepreneurship is considered as the most common powerful economic force across the globe and as a result the inability of many young people in the globe to access paid employment during and after they have graduated calls for the need for entrepreneurship education and training because of its capacity to enable business start-up and management (Keat et al. 2011).

The conception of university entrepreneurship emerged in a similar way as teaching or research. It asserts in the basic principle that ideas from different field of knowledge (e..g, science, arts, engineering, laws, etc.) should be converted into action. In this way, the role of the university is to provide an infrastructure and a culture that allows the ideas to grow (Toone, 2016).

Entrepreneurship education has been defined in various ways by various scholars and practitioners. However, these definitions share a high level of similarity among them. Isaacs et al. (2007) define entrepreneurship education as the purposeful intervention that is made by an educator in the life of the learner through entrepreneurial qualities and skills teaching, which will enable the learner to survive the dynamics of the business world. Mwangi (2011) believes that entrepreneurship education is designed to specifically support graduates, operating and aspirant entrepreneurs in the setting up/operation of their own entrepreneurial ventures rather than to seek paid employment from someone else or institutions (either public/private). Hence, Mensah (2013) adds that entrepreneurship education may capacitate an individual to unleash his/her entrepreneurial potential.

Entrepreneurship education programs are not still in the same level of implementation around the world. It has been part of the curriculum in higher education institutions in North America for over fifty years. In fact, the first graduate course in entrepreneurship was offered at Harvard University in 1948 (Katz, 2003). Today, entrepreneurship courses are offered at most universities across the United States. The demand has been driven by the students themselves, who are eager to take courses ranging from business planning and start-up to entrepreneurial finance and technology management. In Europe, entrepreneurship only substantially began to enter the curriculum in the last ten years, although a handful of institutions started earlier (Twaalfhoven and Wilson, 2004). This is in line with other trends, most notably the growth of the venture capital industry to finance innovative, growth-oriented companies.

Entrepreneurship is not widely recognized as a curricular subject within the university and academia. Instead, it is typically considered more relevant the creation of an institutional culture, practice and policies in a way of developing an entrepreneurial spirit and environment within universities. However, this requires a complete paradigm shift for the entire university, including changing the fundamentals of how the university operates and its role in society.

Wilson (2008) considers that entrepreneurship must be treated as an integral part of a multidisciplinary education process. Students should be encouraged to take courses and engage in projects with students from other disciplines, enabling them to draw upon expertise from across the university – engineering, science, design, liberal arts and business. Additionally, entrepreneurship education must be very closely linked with business practice. For that be possible, professors should have previous experience working with start-ups and alumni of the university should be invited to participate in the classroom to speak to students as well as to teach courses. These courses should be structured to be as experiential as possible, incorporating real-life cases, projects, internships and business plan competitions.

### III. ENTREXPLORER PROJECT

*ENTRExplorer* is a European project with a duration of two years, funded by Leonardo Da Vinci programme, coordinated by The Economic Policies Research Unit from the University of Minho (Portugal), with the participation of EDIT VALUE (Portugal), Sketchpixel (Portugal), Coventry & Warwickshire Chamber of Commerce (United Kingdom), Sterische Volkswirtschaftliche Gesellschaft (Austria) and Bulgarian Development Agency (Bulgaria).

The project developed an online serious game [1] about the various issues associated with entrepreneurship where learners can have access to learning contents that will assist them in the acquisition of entrepreneurial skills. The platform is browser based, allowing a broad use on different devices and operating systems. A screenshot of the front-page offered by the game is depicted in Figure 1.

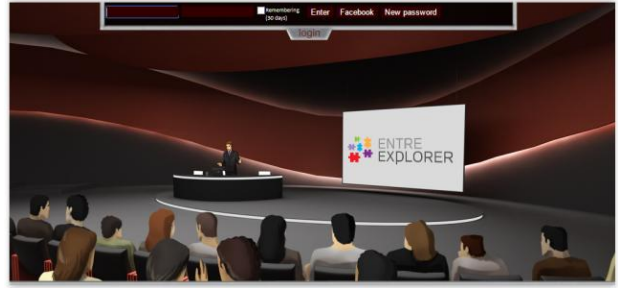

Figure 1.  Front-page of the Entrexplorer game

The gameplay is developed in different scenarios, there is a building with several floors where the player learns about specific subjects and face different challenges. On each floor/level the player has access to the most relevant knowledge about the phases of creating a new business. The information provided in the game guide the player trough the different phases in the process of writing a business plan, he can learn and apply the lessons learned, being able to complete its own business plan, a tool provided in the game.

The player is able to experiment a simulation on the virtual market (Figure 2), which is linked to the previous learning levels, so a good score and performance throughout the learning levels, gives more chance of being successful. A typical badly prepared entrepreneur will have a lower probability of being successful, in the contrary, a well prepared and informed entrepreneur will have a better chance of being successful in the business.

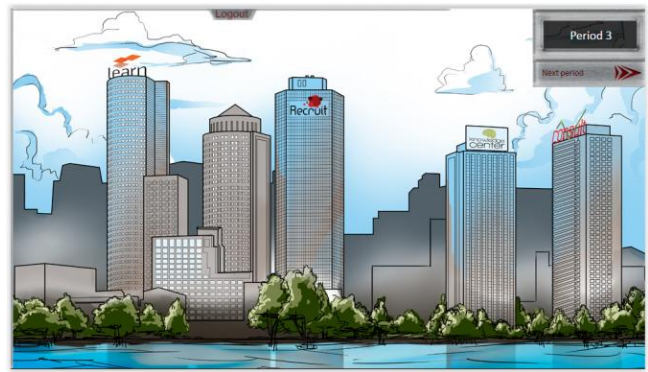

Figure 2.  Virtual market window

---

[1] http://www.entrexplorer.com/game/

The *Entrexplorer* game has the following curriculum units along the eight levels which will be presented and described in the next sections. Each level ends by presenting an assessment questionnaire. There are also six additional levels/floors that offer the following features:

- Creation of an online business plan organized in the same categories as offered by the game;
- Recreation floor where users can watch entrepreneurship videos, play some leisure games, get a list of useful contacts when starting a business and several papers and chapter related to areas such as entrepreneurship, marketing, taxes and finance;
- Lift station where users can fill the entrepreneur profile questionnaire and check the answers given during the game;
- Virtual market that lets the user to simulate a business in a virtual market. The virtual market is linked to the previous eight learning levels, so if the user had a good score and performance throughout the learning levels, then he has more chance of being successful;
- Chat that offers the possibility to interact with other online users;
- Top list where it is possible to check the 15 best successful players of the game and their associated scores.

*A. Level 1: Market and Ideas*

In the first level students will learn how to identify and classify the economic activities by sector (primary sector, secondary sector and tertiary sector. It is given a brief overview about the activities that composes each activity sector. There is also introduced two concepts: workforce and activity rate.

The level introduces some techniques for ideas generation and reveals the importance of creativity in this process. The techniques for ideas generation referred are: brainstorming, brainwriting and SCAMPER.

Understand the key audience is another topic addressed in level 1. For that business value propositions must be relevant to the target market and the target audience should be properly defined, knowing who will buy the product, their age, gender, job, income and so on.

*B. Level 2: Strategic Positioning*

In second level students will learn to understand what it is the strategic positioning. There are three aspects to determine the distinctive features upon which the positioning will be based, namely customers' expectation, the current position of competitors and the benefits of the new product or service. This conceptual triangle proposed by Kotler (2003) is depicted in Figure 3.

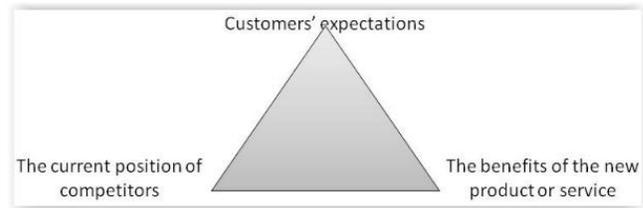

Figure 3. Strategic positioning (Kotler, 2003)

The importance of the corporate mission, vision and values statements is also seen in this chapter. These elements are compared and some examples regarding each concept are given.

*C. Level 3: Product Strategy*

In third level students will learn how to identify and classify products by category. The level begins by making clear the concepts associated with the creation of new products and new services. Then, it asks the user to organize product items in categories such as convenience products, shopping goods and specialty goods.

Within this level it is also given importance to the four paths that companies can adopt to realize what they intend to sell. For that, the game adopts the approach proposed by Kotler (2003) and organize these paths in four categories:

- Selling something that already exists;
- Making something that someone asks for;
- Anticipating something that someone will ask for;
- Making something that no one asked for but that will give buyers great delight.

In the context of this level the user can fill the Entrepreneur questionnaire profile, which is an online self assessment tool, developed within the European project *Your Future - Your Profit* that offers the possibility to get feedback for individuals about their own competences and qualifications regarding the needs and necessities in entrepreneurship, divided into six areas: personal trades, achievement motivation, attitude, framework conditions, skills, knowledge and work experience.

*D. Level 4: Price Strategy*

In level 4 students will learn how to understand the main price strategies. The game establishes that before pricing a product there are crucial aspects that should be taken into account, respectively the production's cost, profit, demand and market competition. Then the game introduces the main pricing strategies that are grouped in the following categories: penetration pricing, skimming pricing, competitive pricing, bundle pricing, product line pricing, premium pricing, cost basing pricing, psychological pricing and optional pricing.

*E. Level 5: Distribution Strategy*

In level 5 students will learn how to recognize the importance of the distribution channels. For that the game starts by clarifying the concept of distribution channel and divides it in direct and indirect channels.

This level also looks to strategies of distribution and groups them in three categories as described below:

- Intensive distribution - it aims to provide saturation of the market by using all available outlets. It is used to distribute low priced or impulsive purchased products, like cigarettes, snack food or soft drinks;
- Selective distribution - a company use a limited or small number of outlets in a certain geographical area to sell her products. This strategy is common when the products are, for example, computers and household appliances;
- Exclusive distribution - only one wholesaler or distribution is used in a specific geographical area. Distribution is limited to a single outlet. Typically this strategy is applied to higher priced products.

Finally, the chapter ends by looking for some problems in establishing an effective distribution strategy. Among them it is relevant to highlight the reluctance to establish different distribution channels for different products, failure to periodically reconsider and update distribution strategies and lack of creativity and resistance to change.

*F. Level 6: Communication Strategy*

In level 6 students will learn to explore the main objectives to achieve in a communication strategy. The game states that whether the communication strategy is designed for a specific project or organizational strategy along time, it should establish the following elements: objectives, audience, messages, tools and activities, resources, timescales, and evaluation and amendment.

This level also explains how the communication process works. For that it presents the traditional communication cycle composed of five stages: sender, encoding, message, decoder and receiver.

*G. Level 7: SWOT Analysis*

In level 7 students will learn to understand and make a SWOT analysis. The four categories of SWOT analysis are presented and detailed.

The idea of SWOT analysis is to evaluate, through an in-depth reflection in which all top managers of the company should participate, what are these four elements. For that, previously we need to gather information about internal (e.g., strengths and weaknesses) and external (e.g., opportunities and threats). Finally, it is necessary to construct a framework with these four elements: on one side the strengths and weaknesses and on the other the opportunities and threats.

In this level, the game offers a mini game where users can put several propositions within the correct group (strength, weakness, opportunity or threat).

*H. Level 8: Financial Viability*

In level 8 students will learn/know the main financial maps. The financial viability analysis is an essential tool that supports entrepreneurs in the decision of going ahead with their business idea. This analysis is based on future economic forecasts and intends to assess the ability of a business to generate sufficient income to meet operating payments, debt commitments and, in some cases, to stimulate the company's growth while maintaining service levels.

The complexity of an economic and financial study depends upon the investment required and the company size. Anyway, some basic concepts in the viability assessment process includes looking for sales, costs of goods sold (COGS), selling, general and administrative expenses (SGA), investments and funding.

Finally, this level presents two elementary financial maps: balance sheet and profit & loss statement. The balance sheet presents the concepts of asset, liability and shareholder equity. For the profit & loss statement, it has introduced the concepts of gross margin, operating profit earnings before interest and taxes (EBIT), earnings before interest and taxes, depreciation and amortization (EBITDA), income before taxes and net income.

## IV. CONCLUSIONS

Serious games offer the potential of improving learning processes by providing attractive, motivating and effective tools that may also create positive situations among students and with teachers. Serious games are the ideal model to match the content in different ways and incorporate problem-solving and reasoning stimulation, which results in learning core competencies. In entrepreneurship field these elements are particularly relevant because it involved multidisciplinary skills, creative, exploratory and argumentative ways of thinking.

The ENTRExplorer game offers an online serious game that intends to promote the development of entrepreneurial skills and help organizations to work smarter by providing a broad understanding of business and organizational dynamics. Stimulating innovative ideas and turning them into value-creating profitable business activities, drive and prepare young people and higher education students to set up their own occupation and create new jobs.

Currently the game has been used in the context of the curricular unit of Technologies and Business, which is placed in the 3rd curricular year of a university. The curricular unit is incorporated in the context of the courses of Management, Informatics and Industrial Management. The game has been adopted by students in the learning process of the concepts of entrepreneurship, innovation and management. Its adoption has helped students in acquiring theoretical knowledge in those domains, but also it has been useful in their work group projects, where they are asked to create the business plan of a fictitious company in the field of information technologies.